\documentclass[12pt]{iopart}
\usepackage{graphicx}
\usepackage{grffile}
\usepackage{iopams}
\expandafter\let\csname equation*\endcsname\relax
\expandafter\let\csname endequation*\endcsname\relax
\usepackage{amsmath}
\usepackage{amssymb}
\usepackage{color}
\usepackage{tabularx}
\usepackage{multirow}  
\usepackage{comment}
\usepackage{url}
\usepackage{txfonts}
\usepackage[ddmmyyyy,hhmmss]{datetime}
\usepackage{anyfontsize}
\newcolumntype{C}{>{\arraybackslash}p{7em}}

\def\nabix{N_{_{\!AB}}^{(ij)}}
\def\nai{N_{_{\!A}}^{(i)}}
\def\nbx{\tilde{N}_{_{\!B}}^{(j)}}
\def\pai{P_{_{\!A}}^{(i)}}
\def\pbx{P_{_{\!B}}^{(j)}}
\def\cabix{c_{_{\!AB}}^{(ij)}}
\def\gix{O^{(ij)}}
\def\nix{N^{(ij)}}
\def\gox{\gamma_{\mathrm{opt}}^{(j)}}

\begin{document}
\title[Generalized gravity model for human migration]{Generalized gravity model for human migration}
\author{Hye Jin Park$^1$, Woo Seong Jo$^{2,3,4}$, Sang Hoon Lee$^{5,\ast}$, and Beom Jun Kim$^{2,\dagger}$}

\address{$^1$Department of Evolutionary Theory, Max Planck Institute for Evolutionary Biology, 24306 Pl\"{o}n, Germany}
\address{$^2$ Department of Physics, Sungkyunkwan University, Suwon 16419, Korea}
\address{$^3$ Northwestern Institute on Complex Systems (NICO), Evanston, Illinois 60208, USA}
\address{$^4$ Kellogg School of Management, Northwestern University, Evanston, Illinois 60208, USA }
\address{$^5$ Department of Liberal Arts, Gyeongnam National University of Science and Technology, Jinju 52725, Korea}

\ead{${}^\ast$lshlj82@gntech.ac.kr}
\ead{${}^\dagger$beomjun@skku.edu}
\vspace{10pt}

\begin{abstract}
The gravity model (GM) analogous to Newton's law of universal gravitation has successfully described the flow between different spatial regions, such as human migration, traffic flows, international economic trades, etc.
This simple but powerful approach relies only on the `mass' factor represented by the scale of the regions and the `geometrical' factor represented by the geographical distance. 
However, when the population has a subpopulation structure distinguished by different attributes, the estimation of the flow solely from the coarse-grained geographical factors in the GM causes the loss of differential geographical information for each attribute.
To exploit the full information contained in the geographical information of subpopulation structure, we generalize the GM for population flow by explicitly harnessing the subpopulation properties characterized by both attributes and geography.
As a concrete example, we examine the marriage patterns between the bride and the groom clans of Korea in the past.
By exploiting more refined geographical and clan information, our generalized GM
properly describes the real data, a part of which could not be explained by the conventional GM. 
Therefore, we would like to emphasize the necessity of using our generalized version of the GM, when the information on such nongeographical subpopulation structures is available.
\end{abstract}
%\keywords{Complex systems, Social systems, Collective behavior}

%\submitto{\jpg}
\maketitle

%\ioptwocol
\section{Introduction}
\label{sec:intro}

For decades, the gravity model (GM) has successfully explained flows between geographically separated two regions such as
traffic flow~\cite{reilly:book:1931,stewart:soc:1948,carrothers:jaip:1956,erlander:book:1990,rodrigue:book:2009}, international economic trades~\cite{tinbergen:book:1962,susan:book:2009}, and human migration~\cite{ravenstein:jssl:1885,anderson:nber:2010}.
The GM is named after Newton's law of universal gravitation because of the similarity in the formula:
a certain type of flow between two regions is proportional to the product of `mass' of each region 
and inversely proportional to a certain power of distance between the regions.
We interpret the mass depending on contexts; we can quantify the relative importance of regions in human migration from their population sizes, and the relative importance of countries in international trades from their economic scales. 
This simple but powerful model has succeeded in interpreting the real world. 
For instance, the GM indeed accurately describes the empirical data of daily human mobility in multiscale mobility networks~\cite{balcan:pnas:2009}.
It also nicely explains the inter- and intra-city traffic flows in Korea~\cite{jung:epl:2008}, along with the passenger flows in the Korean subway system~\cite{goh:pre:2012}.

However, those examples only concern the spatial aspect of population. We can easily imagine more complicated situations such as the population flow between other attributes than the spatial or regional attributes, when the population at a given region consists of \emph{subpopulation} structures. The subpopulation structures characterized by \emph{attributes} for population flows can be ethnic groups, income levels, etc.
In particular, when spatial movement of the subpopulation belonging to one attribute to another subpopulation takes place, the flow between attributes becomes relevant. 
This applies not only to the population flow but also to the international trade, for instance. Goods are transferred in different economic sectors as the attributes, and one may aim at estimating the flow between economic sectors.
One such previous attempt to apply the GM to explain flow between attributes is \cite{lee:prx:2014} partly by a subset of the authors of this paper. 
However, the results have revealed the limitation of the conventional GM when the geographical location of population center of a clan is not representative.
The limitation stems from the process of significant coarse graining of the detailed geographical information of clans into a single point (the population center).

\begin{figure*}[ht]
\centering
\includegraphics[width=1\textwidth]{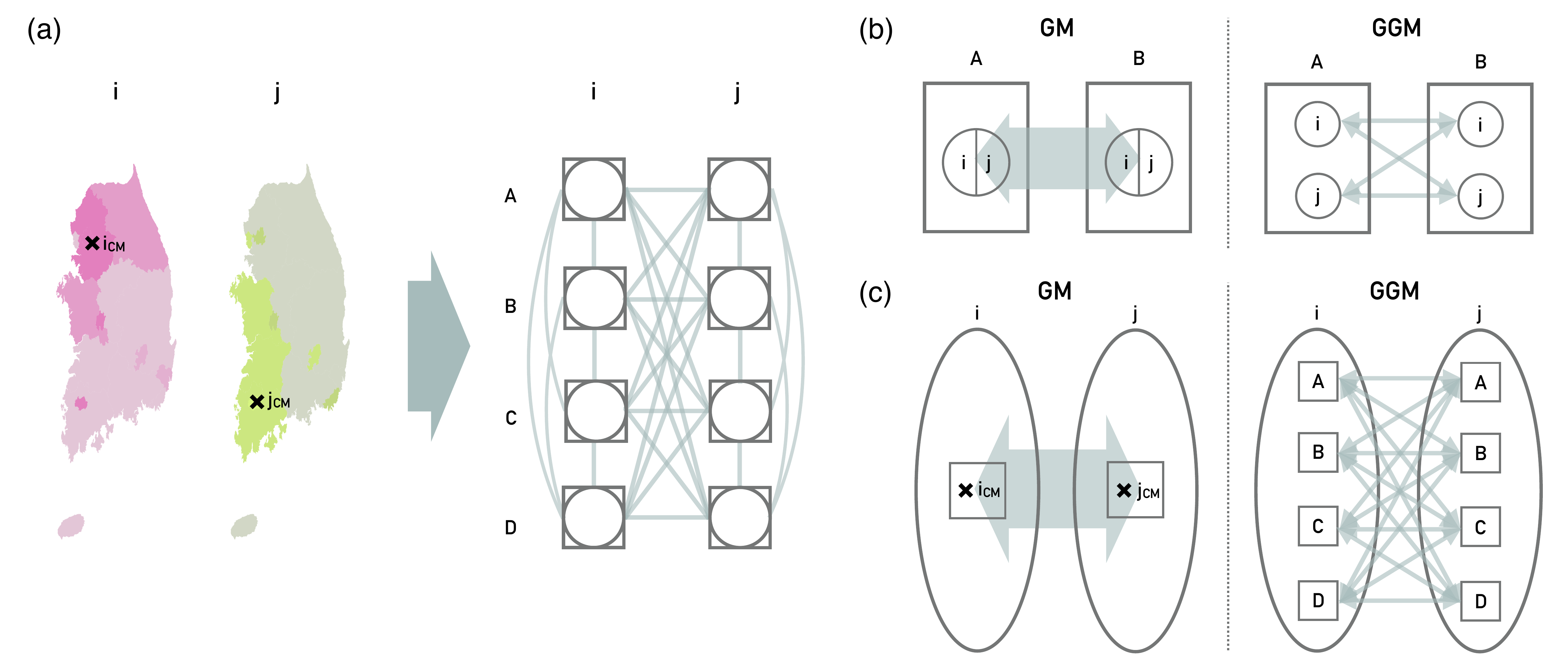}
\caption{
(a) We present the schematic figure for two spatial distributions of attributes, $i$ and $j$, and flows between subpopulations.
Each attribute has its own spatial population distribution represented by the color gradient with the center of mass, e.g., $i_{CM}$ or $j_{CM}$ (marked by the $\times$ symbol). 
Since the population for a given region is again distinguished by its attribute, each subpopulation is distinguished by both region and attribute indices denoted by uppercase letters, $A$, $B$, $C$, and $D$, and lowercase letters, $i$ and $j$, respectively.
We illustrate all of the possible flow combinations between subpopulations,
and we use the rectangular and oval boundaries to distinguish regions and attributes, respectively.  
(b) We show 
the comparison of flows between regions in the viewpoints of GM and GGM.
The GM described on the left considers the flow between regions.
It integrates the population size at a given region first to calculate the flow, while the GGM on the right first considers subpopulation flows and integrates them.
As we show in section~\ref{sec:model}, the results from the GM and GGM are not equivalent.
(c) We compare the flows between attributes in the GM and GGM.
The GM on the left takes into account the flow between attributes $i$ and $j$ from the centers of populations as in \cite{lee:prx:2014},
which causes some information loss due to the coarse graining.
In contrast, we keep the entire information available by integrating all of the subpopulation flows in the GGM (this paper).
}
\label{f.scheme}
\end{figure*}

In this paper, we exploit detailed information on the population substructures by generalizing the GM, rather than coarse graining the information.
First, we formulate the \emph{generalized gravity model} (GGM) to take the full information available on the subpopulation (characterized by both geographical information and attribute information) flows. 
As we show in figure~\ref{f.scheme}(a), when subpopulations are distinguished not only by geographical regions but also by attributes, flows between regions or attributes can be calculated by properly taking subpopulation flows without information loss (see figures~\ref{f.scheme}(b) and (c), respectively).
We also would like to emphasize the necessity of calculating subpopulation flows, because the population flows calculated from the coarse grained population data and those from individual subpopulation data are not equivalent.
As a concrete example, we apply the GGM to marriage records combined with census data. The results show that it effectively captures the geographical constraint imposed in the marriage patterns in the past, in contrast to the GM.

The paper is organized as follows. We first formulate the GGM in section~\ref{sec:model}.
We introduce our data set in section~\ref{sec:data} and apply the model to this data.
The result in section~\ref{sec:res} demonstrates that the GGM indeed captures the geographical information not available from the GM.
Finally, we conclude our work in section~\ref{sec:con}.

\section{The GGM}
\label{sec:model}

Our derivation of the GGM is a natural extension of using the maximum entropy principle to derive the GM~\cite{senior:pg:1979, hua:irsr:1979}, where we replace regional indices with both regional and attribute indices. We provide our step-by-step derivation in this section partly for a pedagogical purpose and the self-containedness of this paper, but most importantly, we can directly demonstrate the problem of using the coarse-grained population data during the derivation. 
The maximum entropy principle is the way to estimate a real probability distribution by maximizing entropy. 
In particular, this method is useful for systems with many degrees of freedom because it focuses on only a few macroscopic quantities.
The real probability distribution is estimated by the maximum entropy principle from the agreement of those observed quantities.
Each Lagrange multiplier corresponding to each constraint in maximization gives the corresponding model parameter.

Let us start from the number $\nabix$ of people who move from region $A$ with attribute $i$ to region $B$ with attribute $j$, where
we take the convention of uppercase letters as the subscript for the region indices, and the lowercase letters as the superscript for the attribute indices.
The sets of attributes $\{i\}$ for the sender side and $\{j\}$ for the receiver side can be different. 
For example, $\{i\}$ and $\{j\}$ can be the education level and the income level, respectively, when we try to describe the flow of employment from one city's education system to another city's industry.
Our formalism adopts the discrete indices and summation, but it should be straightforward to deal with the continuous cases by using continuous variables and integration.

The total number $\nai$ of people moving from region 
$A$ with attribute $i$ to anywhere with any attribute
is then
	\begin{equation}
	\nai = \sum_{j,B} \nabix \,.
	\label{eq:fix}
	\end{equation}
In the same way, the total number $\nbx$ of people who arrive at region $B$ with attribute $j$ from anywhere with any attribute
is given by
	\begin{equation}
	\nbx = \sum_{i,A} \nabix \,.
	\label{eq:fjy}
	\end{equation}
When people move, they have to pay the cost, which is naturally
a function of the
distance between two regions, among other factors.
We denote the cost to move from ($i$, $A$) to ($j$, $B$) for each unit of movement by $\cabix$.
Then, the total moving cost $C$ is the following weighted sum,
	\begin{equation}
	C = \sum_{i,A,j,B} \nabix \cabix \,.
	\label{eq:C}
	\end{equation}

We can then write down the number $W$ of all possible arrangements of travelers considering the multiplicity factor $\nabix$ as
	\begin{equation}
	W = \frac{N!}{\displaystyle \prod_{i,A,j,B}\nabix} \,,
	\label{eq:s}
	\end{equation}
where $N$ is the total number of moving people, $N=\sum_{i,A,j,B} \nabix$.
In the entropy maximization scheme, $\{\nabix\}$ is estimated from 
maximizing the Boltzmann entropy $k_B \log W$ (equivalently maximizing $W$) with constraints. 
We consider three constraints under which $W$ is maximized: the outflows $\{\nai\}$, the inflows $\{\nbx\}$,
and the total moving cost $C$ represented in equations~\eqref{eq:fix}--\eqref{eq:C}.
For this optimization problem under given constraints, we have to use the Lagrange multiplier method, i.e., to find the stationary point of the Lagrangian
	{\footnotesize
	%\begin{widetext}
	\begin{equation}
	\begin{aligned}
	\mathcal{L}(\{\nabix\})&=\frac{N!}{\prod_{i,A,j,B}\nabix} 
		+ \sum_{i,A} \lambda_A^i \left[\nai - \sum_{j,B} \nabix \right]
                + \sum_{j,B} \tilde{\lambda}_B^j \left[\nbx - \sum_{i,A} \nabix \right]
                 + \gamma \left[C - \sum_{i,A,j,B} \nabix \cabix \right] \,,
         \label{eq:L}
	\end{aligned}
	\end{equation}
	%\end{widetext}
	}%
where $\lambda_A^i$, $\tilde{\lambda}_B^j$, and $\gamma$ are 
the Lagrange multipliers for each constraint. 
This problem is essentially the recap of the derivation for the most probable distribution in terms of a given energy value,
from the standard formalism of canonical ensemble in statistical mechanics, so the readers may check the details
in any standard statistical mechanics textbooks such as \cite{huang1987statistical}. The moving cost for each movement and the $\gamma$ parameter here play the roles of energy and inverse temperature there, respectively.
The solution of maximizing equation~\eqref{eq:L} is given by
	\begin{equation}
	\nabix \propto \nai \nbx e^{-\gamma \cabix}.
	\label{eq:fijxy}
	\end{equation}
Note that all $\lambda_A^i$ and $\tilde{\lambda}_B^j$ become unity from the constraint itself (the mass conservation),
so there is only one free parameter $\gamma$, which is determined by the real data. 
Later, we will specifically choose the $\gamma$ value that minimizes the error between the model and real data.

The flow from region $A$ to region $B$ usually decays as a function of the distance between them due to the obviously rising cost,
and thus we have to choose the cost function $c(r)$ as an increasing function of 
the distance $r$ between two regions. 
Conventionally, we set the form $c(r) \propto \ln r$~\cite{chen:chaos:2015}, which leads equation~\eqref{eq:fijxy} to
	\begin{equation}
	\nabix \propto \frac{\nai \nbx}{(r_{AB})^\gamma} \, .
	\label{eq:fijxy2}
	\end{equation}
Note that, though the numbers of leaving or arriving people, $\nai$ and $\nbx$, are not the population size $P_{\text{~region}}^{\text{~attribute}}$ at each region and attribute, generally those are assumed to be linear to the population sizes [$\nai \propto\pai$ and $\nbx\propto\pbx$], yielding 
	\begin{equation}
	\nabix \propto \frac{\pai \pbx}{(r_{AB})^\gamma} \, .
	\label{eq:fijxy3}
	\end{equation}
	%
%[$\nai \propto\pai$ and $\nbx\propto\pbx$]. 
In this case, we can reproduce the GM for the flow between two spatial regions
	\begin{equation}		
	\begin{aligned}
		 N_{AB}\equiv \sum_{i,j}\nabix &\propto \frac{\displaystyle \sum_i \pai \sum_j \pbx}{(r_{AB})^{\gamma}} = \frac{P_A {P}_B}{(r_{AB})^{\gamma}} \,,
	\end{aligned}
	\end{equation}
where $P_A$ and ${P}_B$ are total numbers of people who live in region $A$ and region $B$, respectively.
However, when $\nai$ and $\nbx$ are nonlinear with respect to the population sizes such as in \cite{balcan:pnas:2009}, i.e.,
$\nai \propto [\pai]^{\alpha}$ and $\nbx \propto [\pbx]^{\beta}$,
the population flow from region $A$ with attribute $i$ to region $B$ with attribute $j$: $\nabix \propto [\pai]^{\alpha} [\pbx]^{\beta} / (r_{AB})^\gamma$. In this case, the flow from region $A$ to region $B$ regardless of attributes: $N_{AB}$ cannot be explained by the GM unless $\alpha=\beta=1$ or the population contains only one attribute, because
	\begin{equation}
	\begin{aligned}
		 N_{AB}\equiv \sum_{i,j}\nabix &\propto \frac{\displaystyle \sum_i [\pai]^\alpha \sum_j [\pbx]^\beta}{(r_{AB})^{\gamma}} 
		 & \neq \frac{[P_A]^\alpha [{P}_B]^\beta}{(r_{AB})^{\gamma}} \,.
	\end{aligned}
	\end{equation}
Note that the conventional GM provides the upper or lower bounds: $\sum_i [\pai]^\alpha \leq  [P]^\alpha$ for $\alpha>1$ and $\sum_i [\pai]^\alpha \geq  [P]^\alpha$ for $\alpha<1$, by using the convexity or concavity of the functional form.

In parallel, summing up $\nabix$ for all of the regions gives the
number of moving people from attribute $i$ to $j$,
	\begin{equation}	
	\begin{aligned}
	N^{(ij)} \equiv \sum_{A,B} \nabix &\propto \sum_{A,B} \frac{ \nai\nbx}{(r_{AB})^{\gamma}} .
	\label{e.nix}
	\end{aligned}
	\end{equation}
This is not reducible to the GM either, because 
\begin{equation}
\sum_{A,B} \frac{ \nai\nbx}{(r_{AB})^{\gamma}} \ne \frac{1}{\left[ r(i_\mathrm{CM},j_\mathrm{CM}) \right]^{\gamma}} \displaystyle \sum_{A,B} \nai \nbx \,,
\label{eq:not_proportional_to}
\end{equation}
where $r(i_\mathrm{CM},j_\mathrm{CM})$ is the distance between the centers of population of $i$ and $j$, as shown in figure~\ref{f.scheme}.
The only case when the two expressions actually coincide is the assumption implicitly made in \cite{lee:prx:2014}---the population of each attribute $i$ is treated as the `point mass' located in a single location in space, namely, $i_\mathrm{CM}$.
The GGM estimates the flow between attributes 
without such a coarse graining process involving the information loss.
Hence, the GGM is the correct way to handle subpopulation structures when it comes to the GM of population flow.
We later show that it indeed effectively captures the geographical constraints for the marriage flow in the past obtained from the data,
which was not possible with the coarse-grained version of the GM due to the widely distributed population of the clans~\cite{lee:prx:2014}.

\section{Data sets}
\label{sec:data}

In the traditionally patriarchal culture of Korea after around 17th century, a bride usually moved to her groom's place in the past, once they got married. 
We treat this type of migration caused by the marriage as our main data of human migration.
By applying the GGM to marriage patterns between clans in the past,
we estimate the geographical constraints.
We take the real marriage flow $O^{(ij)}$ from the bride clan $i$ to the groom clan $j$
from the family book data called \textit{jokbo}.
We present more details in section~\ref{sec:jokbo_data_sets}.
To compute the model flow, we extract the distance $r_{AB}$ between two regions and the distribution of the population for each clan, $\nai$ (the bride side) and $\nbx$ (the groom side), from the modern census data in 1985, 2000, and 2015---the three particular years when the information on the regional distribution of each clan's population is available.

We measure the distance $r_{AB}$ based on geographical coordinates of the regions using
the Google maps application programming interface~\cite{gmapi}.
The traveling distance within the same region $r_{AA}$ is estimated as the square root of the region's area. 
We assume that the size of the moving populations, represented by $\nai$ and $\nbx$, are proportional to that of the resident populations (from the census data) of the corresponding clans living in the corresponding regions, so we just take the face values of populations in the census data and treat them as the migrating population for simplicity. As argued in \cite{lee:prx:2014}, we use this modern population data to 
estimate the past migration flow between clans in \textit{jokbo} data, based on the fact that the proportion of each clan living in each region with respect to the total population of Korea has been relatively steady.

\subsection{\textit{Jokbo} data}
\label{sec:jokbo_data_sets}

\begin{table*}[]
\caption{
The volume of the \textit{jokbo} and the census data. 
The number of bride clans and the total number of entries in each \textit{jokbo} are counted based on the existing clans in the census data. 
Note that we count all brides in the \textit{jokbo} whether it includes birth and death dates or not, and thus the volume can be different 
from the previous research~\cite{lee:prx:2014,kiet:jkps:2007,baek:pre:2007,baek:njp:2011}.  
In addition, the population sizes of the groom clans in the census data are presented. 
}
\vspace{5pt}
%\begin{ruledtabular}
{\footnotesize
\begin{tabular*}{1\textwidth}{@{\extracolsep{\fill} } l rr rrr }
 \hline\hline
 \multirow{2}{*}{\textit{jokbo} clan} & \multicolumn{2}{c}{\textit{jokbo} data}   & \multicolumn{3}{c}{census data (population size)}        \\
\cline{2-3} \cline{4-6}
{}   &number of bride clans & number of entries  & year 1985 & year 2000  & year 2015\\   
\hline
1  &  $1\,755$ 	& $155\,392$ & 	$3\,892\,342$	& $4\,324\,478$&$4\,456\,700$ 	\\
2  &  $1\,077$	& $59\,588$  & 	$47\,383$ 		& $61\,650$    &$78\,607$    \\
3  &  $1\,149$  & $54\,377$  &  $200\,334$ 		& $232\,753$   &$298\,092$   \\
4  &   $901$ 	& $25\,542$  & 	$25\,115$ 		& $25\,667$    &$34\,802$    \\
5  &   $782$ 	& $39\,405$  & 	$231\,289$ 		& $238\,505$   &$324\,507$   \\
6 &   $804$ 	& $25\,343$  & 	$21\,756$ 		& $21\,536$    &$27\,343$    \\
7  &  $1\,723$ 	& $189\,158$ & 	$343\,700$		& $380\,530$   &$445\,946$   \\
8  &   $607$ 	& $12\,846$  & 	$15\,539$ 		& $17\,939$    &$20\,484$    \\
9 &   $356$ 	& $5\,146$   & 	$103\,220$ 		& $123\,688$   &$163\,610$   \\
\hline\hline
\end{tabular*}}
\label{t.vol}
\end{table*}

\textit{Jokbo}, or the Korean family book, records the members of paternal lineage and each member's spouse and children.
Even though a bride does not change her family name after marriage in Korean culture, she was (and still is in many conservative families) considered to belong to the groom's family after marriage.
The key element of \textit{jokbo} for our research is the fact that it records information of the female spouse's original clan including the information on its geographical origin.
Each clan has its own \textit{jokbo}, which is passed down to descendants.
Previously, the distributions of clans in Korea have been studied based on ten \textit{jokbo} data~\cite{kiet:jkps:2007,baek:pre:2007,baek:njp:2011}.
Marriage patterns using the same data set have been studied in \cite{lee:prx:2014} 
with the GM framework under the assumption described in the left figure of figure~\ref{f.scheme}(c).

We also use the same ten \textit{jokbo} data set, but at this time we merge two \textit{jokbo} among ten because those two are 
different subgroups of the same clan (because our attribute unit is the clans), which results in the total number of nine distinct \textit{jokbo} used in our analysis.
We count how many brides from clan $i$ married the grooms from clan $j$, the owner of \textit{jokbo},
and treat the number of brides as the real migration flow $O^{(ij)}$ from the bride clan $i$ to the groom clan $j$. 
Each \textit{jokbo} contains between $5\,146$ and $189\,158$ marriage entries (see table~\ref{t.vol} for detailed statistics).
We index the \textit{jokbo} in the ascending order of the value of $\gox$ (that will be introduced in section~\ref{sec:res}) predicted from the 2015 census data.
There is a single case of a tie, and we break it by using $\gox$ from the 1985 data.

\subsection{Census data}
We assume that the outflow $\nai$ from $(i, A)$ and the inflow $\nbx$ to $(j, B)$
are proportional to their population sizes, $\pai$ and $\pbx$, to predict the flow $\nix$ from equation~\eqref{e.nix}.
The population size of each clan residing in each region is taken from
the Korean census data~\cite{kosis1},
where the spatial resolution of the data is determined by the set of $194$ administrative regions.
In particular, we use the census data in the years 1985 and 2000 as in \cite{lee:prx:2014}, and the new data in 2015.
We present the detailed population statistics for each groom clan in the census data in table~\ref{t.vol}.

Due to the changes in the administrative boundaries over 30 years (between 1985 and 2015), 
we have generated the common set of $194$ administrative regions for the three different years,
where we have unified the administrative regions
whose boundaries had been changed, following the procedure of
\cite{lee:prx:2014} to unify the administrative regions (for the two different years: 1985 and 2000, in \cite{lee:prx:2014}).

%%%%%%%%%%%%%%%%%%%%%%%%%%%%%%%%%%%%%%%%%%%%%%%%%%%%%%%%%%%%%%%%%%%%%%%%%%
%------------------------------- results --------------------------------%
\section{Results}
\label{sec:res}

\begin{figure*}
\centering
\begin{tabular}{ccc}
\includegraphics[width=0.33\textwidth]{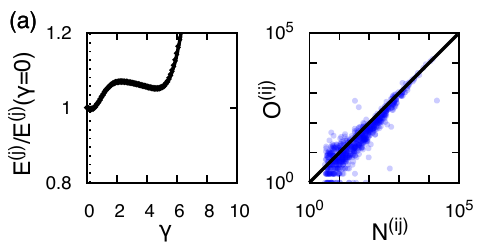}&
\includegraphics[width=0.33\textwidth]{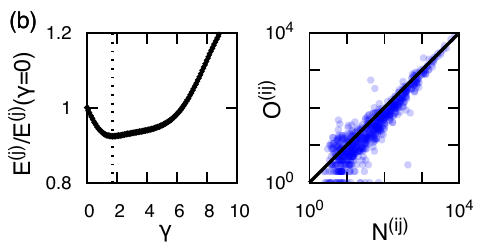}&
\includegraphics[width=0.33\textwidth]{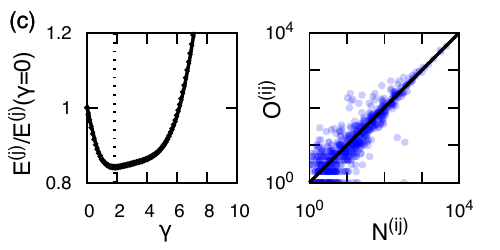}\\
\includegraphics[width=0.33\textwidth]{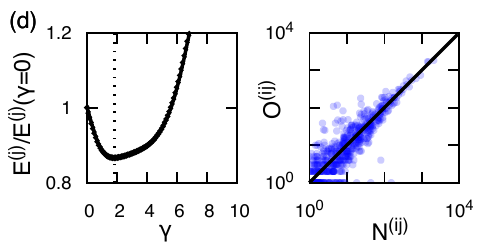}&
\includegraphics[width=0.33\textwidth]{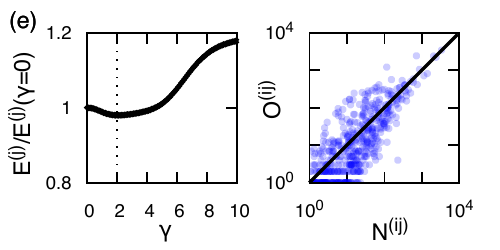}&
\includegraphics[width=0.33\textwidth]{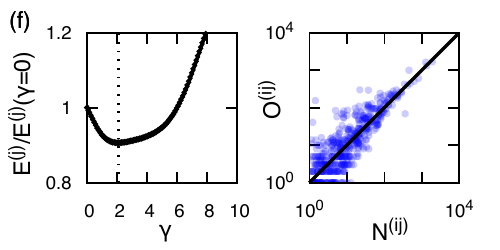}\\
\includegraphics[width=0.33\textwidth]{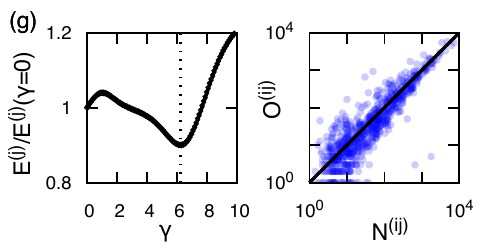}&
\includegraphics[width=0.33\textwidth]{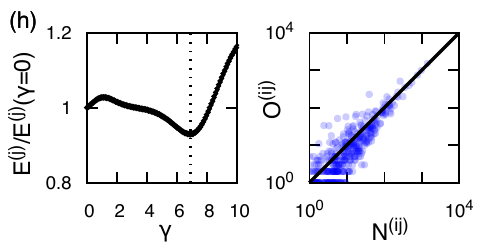}&
\includegraphics[width=0.33\textwidth]{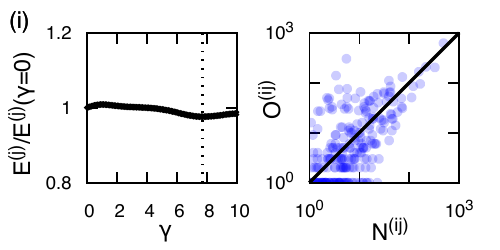}\\
\end{tabular}
\caption{
The error landscapes as the function of $\gamma$ and the scatter plot of real and estimated marriage flows at $\gox$ for each groom clan $j$, with the census data in 2015.
We index the groom clans according to $\gox$ in the ascending order, from $1$ to $9$.
The panels (a)--(i) correspond to the groom clans $1$--$9$, respectively.
For the actual error landscape plots, we use the normalized error $E^{(j)} (\gamma) / E^{(j)}(\gamma = 0)$ with respect to the $\gamma = 0$ case.
The vertical dashed lines in the error landscapes indicate the $\gox$ value that gives the minimum value of $E^{(j)}$. 
}
\label{f.error}
\end{figure*}

We apply the GGM expressed in equation~\eqref{e.nix} to the marriage patterns in the past. 
The real number of marriage entries $\{O^{(ij)}\}$ are listed in the \textit{jokbo} data, and 
we compare the predicted flow $\{\nix\}$ from the model with $\{O^{(ij)}\}$.
For each groom clan (corresponding to each \textit{jokbo} clan) $j$, the difference is quantified by the error 
\begin{equation}
E^{(j)}=\sqrt{\sum_i\left[E^{(ij)}\right]^2}\equiv\sqrt{\sum_{i}\left[\gix-\nix\right]^2} \,,
\label{eq:error_definition}
\end{equation}
calculated from the list of the bride clans $\{ i \}$.
Note that we discard the self migration flow $E^{(jj)}$ when we calculate $E^{(j)}$,
because the marriage between the same clans was forbidden in the past
and it is indeed significantly underrepresented as reported in \cite{lee:prx:2014}. 
In practice, we also checked that ignoring $E^{(jj)}$ does not make much of a difference in our results.
The proportionality factor for equation~\eqref{e.nix} is calculated by minimizing $E^{(j)}$ at a given value of $\gamma^{(j)}$.
The optimal value $\gox$ is assigned as the $\gamma^{(j)}$ value that minimizes the error $E^{(j)}$. In this case, we vary the $\gamma^{(j)}$ value from $0$ to $10$ with the resolution of $0.1$.
The obtained $\gox$ value indicates the geographical constraint for the brides' migration to the groom clan $j$.

In figure~\ref{f.error}, on the left side of each panel (corresponding to each groom clan $j$), we show the error in equation~\eqref{eq:error_definition} as a function of $\gamma^{(j)}$, via the fact that $\nix$ is a function of $\gamma^{(j)}$.
On the right side of each panel, we also show scatter plots comparing the real flow $O^{(ij)}$ versus the predicted flow $N^{(ij)}$ from the GGM at a given $\gox$ value with the guideline corresponding to $\gix = \nix$.
The predicted flow $\nix$ and the number of entries $O^{(ij)}$ in \textit{jokbo} are indeed close to the $\gix = \nix$ line.
Since the results from 1985, 2000, and 2015 census data are qualitatively the same,
we only show the results in 2015. 
Except for the clan $1$, the GGM actually captures nonzero $\gox$ values, while the exponent $\gox$ always vanishes when we use the GM for all of the \textit{jokbo}~\cite{lee:prx:2014}.

\begin{figure*}
\centering
\includegraphics[width=0.6\textwidth]{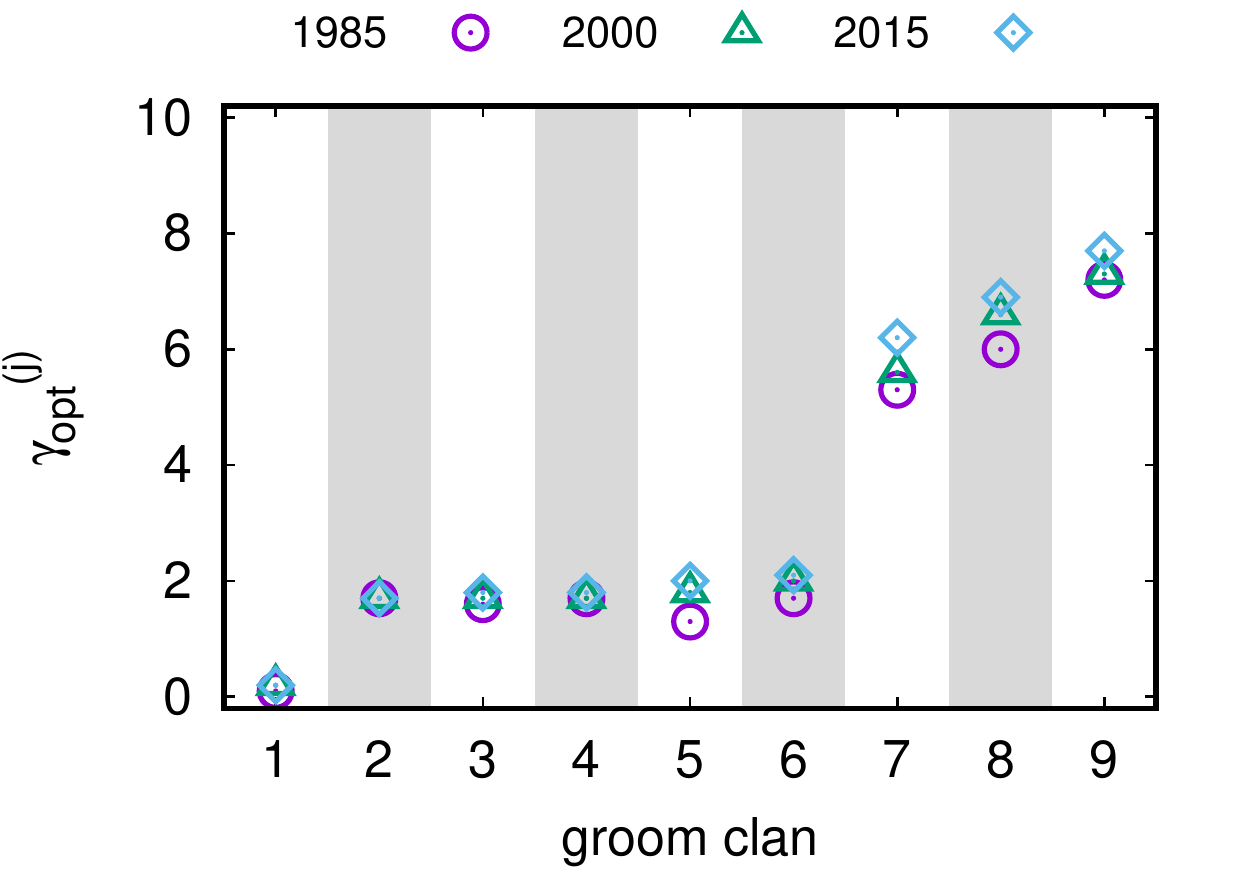}
\caption{The estimated exponent $\gox$ of distance in equation~\eqref{e.nix} from the census data in 1985, 2000, and 2015, for each groom clan $j$. 
The horizontal axis indicates groom clans, and we use three different types of symbols to distinguish the results for each year. 
We shade every other column for better readability.
}
\label{f.gamma}
\end{figure*}

\begin{table*}[]
\caption{The results of $\gox$ for each groom clan $j$, from 1985, 2000, and 2015 census data, respectively.  
For the comparison between the GGM and the GM, 
we provide the values of $\Delta e^{(j)}$ (\%) in equation~\eqref{eq:normalied_error} as the percentage, representing the relative performance of the GGM.
We also characterize the population distribution of each clan by measuring the dispersion $\Delta R^{(j)}$ (km) in equation~\eqref{eq:deltaR} and the effective number of occupied regions $n^{(j)}$ in equation~\eqref{eq:h}. 
}
\vspace{5pt}
\scriptsize
\begin{tabular*}{1\textwidth}{@{\extracolsep{\fill} } l  llll llll llll}
\hline\hline
\multicolumn{1}{c}{\multirow{2}{*}{\shortstack{groom \\clan $j$}}}  & \multicolumn{4}{c}{year 1985}   & \multicolumn{4}{c}{year 2000}  & \multicolumn{4}{c}{year 2015} \\ \cline{2-5} \cline{6-9} \cline{10-13}
\multicolumn{1}{c}{}  & $\gox$ & $\Delta e^{(j)} $ & $\Delta R^{(j)}$ & $n^{(j)}$  & $\gox$ & $\Delta e^{(j)}$ & $\Delta R^{(j)}$ & $n^{(j)}$ & $\gox$ & $\Delta e^{(j)}$ & $\Delta R^{(j)}$ & $n^{(j)}$\\
\hline 
1  & $0.1$	& $0.02$  & $151.35$ & $104.23$	  & $0.2$   & $0.17$   & $148.54$   &  $81.43$  & $0.2$ & $0.38$   & $146.32$   &  $76.65$     \\
2  & $1.7$	& $9.98$  & $118.39$ & $43.15$    & $1.7$   & $8.50$   & $118.43$   &  $43.18$	& $1.7$ & $7.57$   & $117.83$   &  $47.21$    \\
3  & $1.6$	& $17.81$ & $149.91$ & $71.31$    & $1.7$   & $18.99$  & $142.11$   &  $62.44$	& $1.8$ & $16.81$  & $139.23$   &  $60.14$    \\
4  & $1.7$	& $15.01$ & $131.46$ & $85.29$    & $1.7$   & $15.17$  & $125.76$   &  $67.34$	& $1.8$ & $13.51$  & $125.52$   &  $64.48$    \\
5  & $1.3$	& $0.52$  & $143.01$ & $54.98$    & $1.8$   & $1.28$   & $146.06$   &  $51.84$ 	& $2.0$ & $2.02$   & $146.76$   &  $62.44$    \\
6 & $1.7$	& $9.43$  & $141.09$ & $60.74$    & $2.0$   & $10.12$  & $137.44$   &  $50.50$  & $2.1$ & $9.24$   & $134.61$   &  $51.60$   \\
7  & $5.3$	& $6.05$  & $155.77$ & $91.37$    & $5.6$   & $9.69$   & $152.98$   &  $74.21$	& $6.2$ & $10.32$  & $149.64$   &  $71.26$    \\
8  & $6.0$	& $2.88$  & $140.40$ & $63.56$    & $6.6$   & $6.64$   & $136.73$   &  $61.83$	& $6.9$ & $7.08$   & $133.10$   &  $57.67$     \\
9  & $7.2$	& $2.18$  & $139.94$ & $87.32$    & $7.3$   & $2.54$   & $136.68$   &  $70.49$	& $7.7$ & $2.40$   & $134.97$   &  $69.69$    \\
\hline\hline
\end{tabular*}
\label{t.g}
\end{table*}

We calculate $\nix$ from each census data in 1985, 2000, and 2015, so
we obtain the three $\gox$ values corresponding to each year, for each groom clan (see figure~\ref{f.gamma}).
From the similarity of error landscapes in each census data, the $\gox$ values for different years are not much different for a given groom clan $j$ (check table~\ref{t.g} for details), which indicates that the results of $\gox$ are temporally robust for a given groom clan.
It is also interesting to note that many $\gox$ values are around $2$ corresponding to the same formula with demographic gravitation introduced in \cite{stewart:soc:1948}.
Most importantly, compared with the GM results, i.e., $\gox \simeq 0$ for all of the groom clans~\cite{lee:prx:2014},
the GGM indeed yields nonzero $\gox$ values except for the clan $1$. 
It implies that the GGM actually captures the information of the geographical constraint on the flow, in contrast to the GM where it is hard to capture this geographical information due to the coarse graining, i.e.,
treating the population of a clan as a point particle at a center of mass.
On the contrary, the GGM uses much more detailed information of the subpopulation structure that eventually leads us to capture the actual geographical constraint imposed on the flow.

\begin{figure*}
\centering
\includegraphics[width=0.6\textwidth]{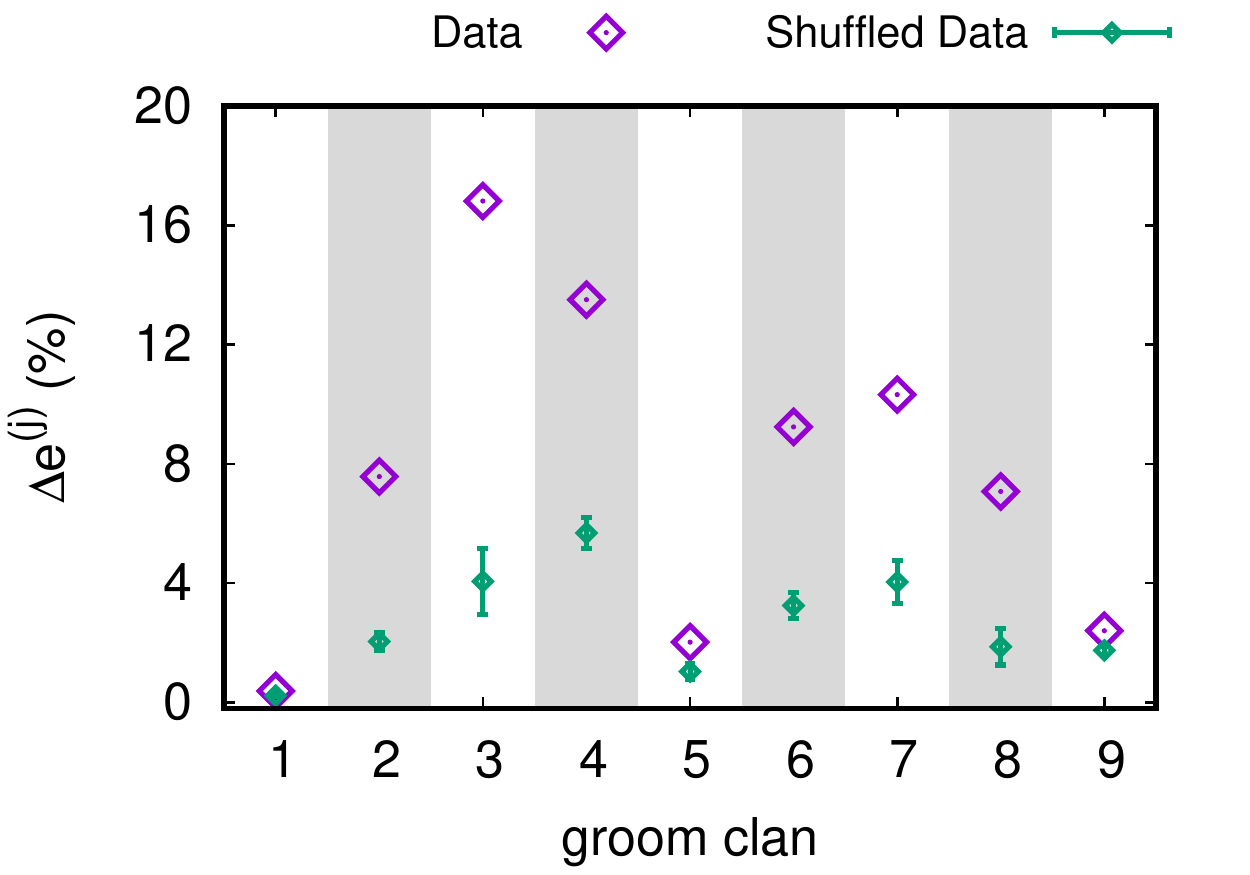}
\caption{The performance of the GGM expressed as the normalized reduced error in equation~\eqref{eq:normalied_error} from the census data in 2015 for each groom clan $j$, compared with the model performance from geographically scrambled data. The values for the real data and the average values from the shuffled data are shown as the large purple diamonds and the small green diamonds with the error bars representing the standard deviation over $100$ realizations, respectively.
As in figure~\ref{f.gamma}, The horizontal axis indicates groom clans and we shade every other column for better readability.
}
\label{f.shuffle}
\end{figure*}

To validate our model, we measure the performance of our model compared with the GM using the normalized reduced error with respect to the case of no geographical constraint, i.e., $\gamma = 0$, defined as
	\begin{equation}
		\Delta e^{(j)} = \frac{\left[ E^{(j)}\left(\gamma=0\right)-E^{(j)}\left(\gamma=\gox\right) \right] }{ E^{(j)}\left(\gamma=0\right)} \,.
	\label{eq:normalied_error}
	\end{equation}
The normalized reduced error $\Delta e^{(j)}$ quantifies the improvement of performance by using the GGM compared with the GM, which results in $\gox \simeq 0$ for all of the clans~\cite{lee:prx:2014}.
Large values of $\Delta e^{(j)}$ indicate significance of geographical constraints in the migration flow.
Except for clans such as $j=$1, 5, and 9, the normalized reduced error $\Delta e^{(j)} \simeq 10\%$, as shown in figure~\ref{f.shuffle}.
To demonstrate the statistical significance of geographical information of the data, we shuffle regional indices for the groom clan $j$ to obtain the corresponding surrogate $\gox$ values, also shown in figure~\ref{f.shuffle}.

For the exceptional cases of $j= 1$, $5$, and $9$, we suspect the lack of geographical information in the data itself, as we argue.
To test the statistical significance of spatial correlation in the data, 
we shuffle the regional indices to scramble geographical information. 
We examine the result of our model in this shuffled data, which is shown in figure~\ref{f.shuffle}.
It supports that the GGM extracts more geographical information than the GM, by capturing the nonzero $\gamma$ exponent.
If the shuffled data gives similar results to those from original data, the original data contains a small amount of geographical information.
As shown in figure~\ref{f.shuffle}, this situation precisely happens for the clans $j=1$, $5$, and $9$, whose original and shuffled data give similar results. In other words, the data corresponding to those three clans originally contain less geographical information than the other clans.
The small $\Delta e^{(j)}$, therefore, does not come from the GGM but from the clan data itself.
Hence, we conclude that as long as the data has enough geographical information, the GGM effectively extracts the corresponding information.

As mentioned above, geographical information of the population distribution is closely related to the model performance and the statistical significance of the nonzero $\gox$ values.
To quantify the geographical information in the distribution of clans more systematically, we introduce two measures:
the dispersion that quantifies how strongly localized the populations are, and 
the homogeneity that focuses on how uniformly the populations occupy distinct regions. 
For the latter, we use the concept of the effective number of occupied regions based on the R{\'e}nyi entropy for a given probability distribution, as in \cite{lee:plosone:2010}.
We define the dispersion $\Delta R^{(j)}$ from the centroid $\mathbf{R}^{(j)}$ of the clan $j$ by taking population fractions as weights: 
	\begin{equation}
		\Delta R^{(j)} = \sqrt{\sum_A f_A^{(j)} \left|\left|\mathbf{r}_A-\mathbf{R}^{(j)}\right|\right|^2},
	\label{eq:deltaR}
	\end{equation}
where $\mathbf{r}_A$ is the location of the administrative region $A$. The population fraction $f^{(j)}_A$ is 
the population of clan $j$ living in $A$ divided by the total population of clan $j$,
and $|| \cdots ||$ is the Euclidean norm.
The population centroid of clan $j$ is then $\mathbf{R}^{(j)}=\sum_A f_A^{(j)}\mathbf{r}_A$.
As the concept of moment of inertia or radius of gyration~\cite{lee:prx:2014}, $\Delta R^{(j)}$ measures how (geographically) widely a certain clan is distributed.
The effective number of occupied regions is defined as the reciprocal of the heterogeneity quantified by the second moment of the population fraction, given by
	\begin{equation}
		n^{(j)}=\frac{1}{\displaystyle \sum_A \left[f_A^{(j)}\right]^2}.
	\label{eq:h}
	\end{equation}
In contrast to $\Delta R^{(j)}$, $n^{(j)}$ measures how many of distinct regions (regardless of their geographical location) a certain clan occupies effectively; there are scaling relations for extreme cases: $n^{(j)} \simeq$ the total number of administrative regions when the clan $j$ is uniformly distributed to the entire set of administrative regions, while $n^{(j)} \simeq 1$ when the clan $j$ is almost exclusively living in a single particular administrative region~\cite{lee:plosone:2010}.

\begin{figure*}
\centering
\includegraphics[width=0.6\textwidth]{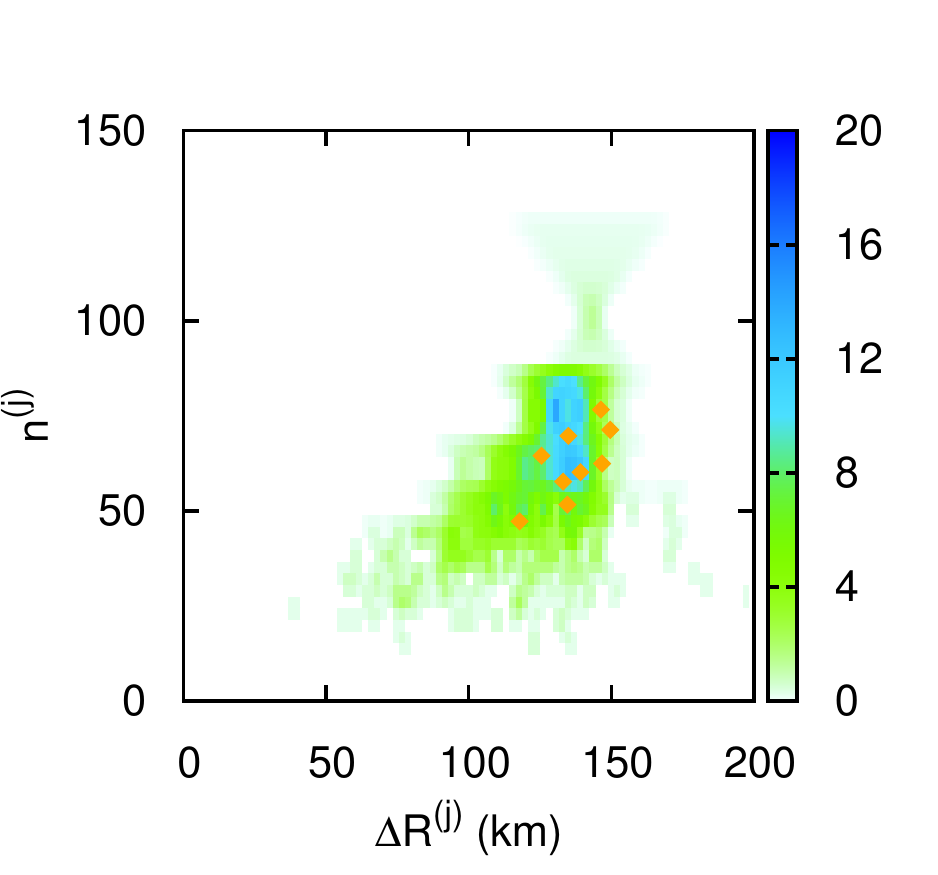}
\caption{The density plot of dispersion $\Delta R^{(j)}$ and the effective number of occupied regions $n^{(j)}$ for all of the clans in the 2015 census data.
The nine orange diamonds correspond to the groom clans in the \textit{jokbo} data. 
}
\label{f.hist}
\end{figure*}

The dispersion $\Delta R^{(j)}$ and the number $n^{(j)}$ of occupied regions are usually positively correlated. 
However, $\Delta R^{(j)}$ can be large even when $n^{(j)}$ is small, e.g., when the clan has multiple localized residential regions.
Hence, we use both measures for more accurate identification of the population distributions.
For all of the $788$ clans in the 2015 census data, we measure both $\Delta R^{(j)}$ and $n^{(j)}$ and 
present them as the density plot in figure~\ref{f.hist}.
Among all of the clans listed in the census data, all of the groom clans corresponding to the \textit{jokbo} data have relevantly large values of dispersion and the effective number of occupied regions.
Note that the combination of large $\Delta R^{(j)}$ and small $n^{(j)}$ (as discussed before) is observed indeed, while the combination of small $\Delta R^{(j)}$ and large $n^{(j)}$ does not appear, as shown in figure~\ref{f.hist}. This contrast hints the existence of such multiple localized residential regions, which is also discussed in \cite{lee:prx:2014}.

\begin{figure*}
\centering
\includegraphics[width=0.6\textwidth]{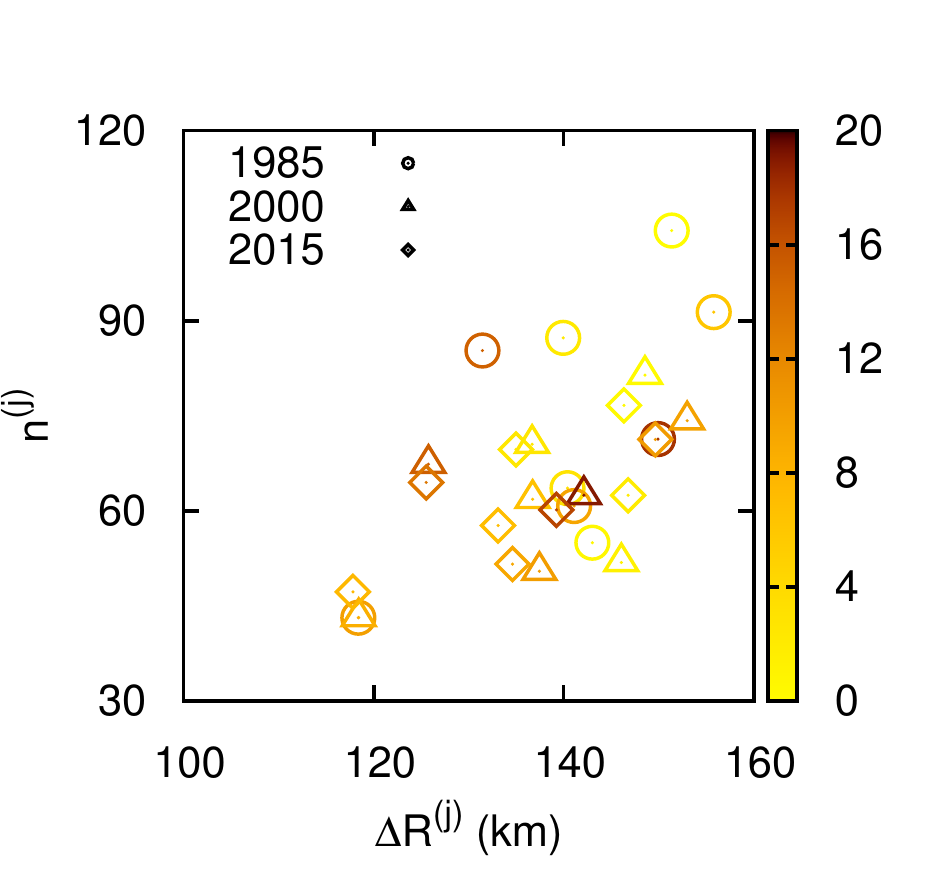}
\caption{The same $\Delta R^{(j)}$--$n^{(j)}$ diagram as figure~\ref{f.hist}, but only with the groom clans, using the census data in the three different years. The color of the points represents $\Delta e^{(j)}$, and the
different symbols indicate the results from the census data in different years. 
}
\label{f.rl}
\end{figure*}

Finally, we compare the relative performance of GGM with these two measures, presented in figure~\ref{f.rl}. 
There is a trend of increasing $\Delta e^{(j)}$ when the population distribution has more geographical information, low dispersion and low effective number of occupation regions. The Pearson correlation coefficients between $\Delta e^{(j)}$ and the dispersion $R^{(j)}$ or the effective number $n^{(j)}$ of occupied regions are $-0.270$ and $-0.213$, respectively, implying the anticorrelation.
This observation again confirms that when the data includes meaningful geographical information, the GGM captures the geographical constraint, while the GM may not.

%%%%%%%%%%%%%%%%%%%%%%%%%%%%%%%%%%%%%%%%%%%%%%%%%%%%%%%%%%%%%%%%%%%%%%%%%%
%------------------------------- conclusion ------------------------------%
\section{Conclusions and discussion}
\label{sec:con}
In this paper, we formulate the GGM to properly take subpopulation structures for human migration, 
by keeping the entire geographical distribution of the subpopulations. 
The key aspect of our point is that we need to calculate individual subpopulation flows, before trying any geographical coarsening.  
To test the validity of the GGM, we investigate the marriage patterns of Korea in the past.
Applying our model to the marriage pattern, we identify the geographical constraint.
The results demonstrate that the GGM captures the subpopulation aspect of the data without the information loss occurred in the GM.

We believe that our approach is applicable to a wide range of research on population dynamics. 
Moreover, we would like to point out that the GGM is in fact even more general than the treatment for our particular data set, e.g., the different types of attributes for the departure and arrival places by taking the different sets $\{i\} \ne \{j\}$.
For instance, the attribute in the departure place for education can be the education level of people,
while the attribute of the arrival place for work can be the income level. 
Furthermore, if we release the constraints $\alpha=1$ and $\beta=1$, one can also allow the nonlinear mass relation. 
In this case, the GGM becomes particularly important to prevent the loss of information because the scale factor, in addition to the distance factor, also has the nonlinear relation with the flow.

Finally, this scheme can be extended for multiple types of attributes that can be represented by the attribute vectors.
For instance, people living in the region $A$ with the attribute $\mathbf{i} = (e_i, m_i)$ representing the education level $e_i$ and the income level $m_i$ can move to the region $B$ with $\mathbf{j} = (e_j, m_j)$.
We hope to extend this type of general scheme for a wide variety of different data sets of human (and possibly nonhuman) migration or flow patterns in the future.

\ack
S.H.L. was supported by Gyeongnam National University of Science and Technology Grant in 2018--2019. B.J.K. was supported by the National Research Foundation of Korea (NRF) grant funded by the Korea government (MSIT) (No. 2017R1A2B2005957). We appreciate the anonymous referee for pointing out that the conventional gravity model plays the role of upper or lower bounds.

%%%%%%%%%%%%%%%%%%%%%%%%%%%%%%%%%%%%%%%%%%%%%%%%%%%%%%%%%%%%%%%%%%%%%%%%%%
%--------------------------------- ref ----------------------------------%
\section*{References}
%\bibliographystyle{iopart-num}
%\bibliography{jokbo}
\providecommand{\newblock}{}

\end{document}